\documentstyle[preprint,aps,epsf]{revtex}

\begin{document}
\tighten
\newcommand{\be}{\begin{equation}}
\newcommand{\ee}{\end{equation}}
\newcommand{\bea}{\begin{eqnarray}}
\newcommand{\eea}{\end{eqnarray}}
\newcommand{\r}{\rangle }
\newcommand{\xx}{\noindent }

\draft

\title{A General Expression for Symmetry Factors of Feynman Diagrams}

\author{ C.D. Palmer${}^{a}$ and M.E. Carrington${}^{b,c}$}

 \address{${}^a$ Department of Mathematics, Brandon University, Brandon, Manitoba, R7A 6A9 Canada \\ 
 ${}^b$ Department of Physics, Brandon University, Brandon, 
Manitoba,
R7A 6A9 Canada\\
 ${}^c$  Winnipeg Institute for Theoretical Physics, Winnipeg, Manitoba, R3B 2E9 Canada}
 
 \date{\today}

\maketitle

\begin{abstract}
The calculation of the symmetry factor corresponding to a given Feynman diagram is well known to be a tedious problem. We have
derived a simple formula for these symmetry factors. Our formula works for any diagram in scalar theory ($\phi^3$ and
$\phi^4$ interactions), spinor QED, scalar QED, or QCD.


\end{abstract}

\narrowtext

\section{Introduction}

In field theoretic calculations, amplitudes are calculated from the generating
functional by performing functional derivatives. The Feynman rules are a handy short cut. One
draws all the topologically distinct diagrams and uses a set of rules, derived from the
generating functional, to represent the diagram as a integral over momentum variables. The Feynman rules
tell us what factor to include for each propagator, vertex and loop momentum.
However, the overall numerical factor has to be determined separately. This factor is called the
symmetry factor of the diagram \cite{bb}. One attempts to choose the 
coefficients of the
interaction terms in the Lagrangian so that the symmetry factor is one. However, for any theory that has an interaction
which contains more than one factor of the same field, there will be diagrams for which the symmetry factor is not 
one.  Spinor QED is an example of a theory for which all symmetry factors are one. The interaction part of the Lagrangian has the form $\bar \psi \gamma_\mu A^\mu \psi$.  In this expression the electron, positron and photon fields appear once each. In contrast, in scalar QED, there is an interaction of the form $\phi^\dagger A_\mu A^\mu \phi$. The presence of two powers of the photon field will result in non-unity symmetry factors for some diagrams.  Similarly, $\phi^3$ theory, $\phi^4$ theory and QCD will have non-unity symmetry factors for some diagrams. 

\section{Symmetry Factors for Scalar Theories}

We will write the symmetry factor as 
\bea
S={\cal N}/{\cal D} \label{S}
\eea
where ${\cal N}$ and ${\cal D}$ stand for numerator and denominator respectively. We describe below how to determine the numerical
values of ${\cal N}$ and ${\cal D}$. For simplicity, we begin with scalar theory. Consider the generating function for connected n-point functions for a scalar theory. We
have,
\bea
W[J] = {\rm exp} \left( \frac{i}{2} \int dx \int dy\, J(x) D(x-y) J(y) + i \int dx {\cal L}_{int}\left[
\frac{1}{i} \frac{\delta}{\delta J(x)}\right]\right)\label{gen1}
\eea
We consider a theory with both cubic and quartic interactions:
\bea
{\cal L}_{int} = \frac{g}{3!}\phi^3 + \frac{\lambda}{4!}\phi^4
\label{lint}
\eea
The factors $3!$ and $4!$ in the expression above are included so that as many diagrams as possible will have a symmetry factor of one. 
We adopt the following short hand notation: $D_{xy}:=D(x-y);$  $J_x:= J(x);$  $DJ_x := D_{xy} J_y;$ and $JDJ := J_x D_{xy} J_y$.
The integrals over the spatial variables will be suppressed. In addition, we will drop all factors of $i$ and not keep track of signs, since we are only interested in the numerical
factor. \\

We look at the calculation of the amplitude corresponding to the diagram in Fig. [1]. 
We
obtain,
\bea
G(w,x,y,z) \simeq \frac{\delta}{\delta J_w}\frac{\delta}{\delta J_x} \frac{\delta}{\delta J_y} \frac{\delta}{\delta J_z} W[J]|_{J=0} 
\eea
where the notation $\simeq$ indicates that signs and factors of $i$ have been dropped. 
We expand the exponential and keep only the terms that give the diagram we are considering. 
We obtain:
\bea
{\cal D}\,\, \frac{\delta}{\delta J_w}\frac{\delta}{\delta J_x} \frac{\delta}{\delta J_y} 
\frac{\delta}{\delta J_z} \left(\frac{\delta}{\delta J_a}\right)^4 \cdot \left(\frac{\delta}{\delta J_b}\right)^4 
\cdot \left(\frac{\delta}{\delta J_c}\right)^4 \cdot \left(JDJ\right)^8
\eea
The factor ${\cal D}$ is given by:
\bea
\frac{1}{{\cal D}}= \frac{1}{3!} \left(\frac{1}{4!}\right)^3\left(\frac{1}{2}\right)^8 \frac{1}{8!}\label{D}
\eea
This expression is obtained as follows:

\xx a) $\frac{1}{3!}$ from expanding ${\rm exp}({\cal L}_{int})=\sum_n \frac{1}{n!}
({\cal
L}_{int})^n$ and taking the third term in the
expansion which contributes three factors of the interaction Lagrangian

\xx b) 3 factors of
$1/4!$ from ${\cal L}_{int}$ (Eqn. (\ref{lint}))

\xx c) $\left(\frac{1}{2}\right)^8 \frac{1}{8!}$ from expanding 
${\rm exp} \left( \frac{i}{2} \int dx \int dy\, J(x) D(x-y) J(y)\right)$ and keeping the term that
contains eight propagators\\

The factor ${\cal N}$ is the number of ways that the functional derivatives can act on the factor $(JDJ)^8$ to produce 
the
combination of propagators that corresponds to the diagram we are looking at. The first step is to include a factor 3! which comes from the possible
permutations of the indices $\{a,b,c\}$. Including this factor allows us to restrict ourselves to the product of progagators:
$D_{ax}D_{aw}D_{ab}(D_{bc})^2 D_{ac}D_{by}D_{cz}$. Now we begin taking derivatives. We start with the derivatives that
correspond to external legs. We obtain,
\bea
16 \cdot 14 \cdot 12 \cdot 10 \cdot DJ_w \cdot DJ_x \cdot DJ_y \cdot DJ_z 
\cdot \left(JDJ \right)^4
\eea
Note that there are no terms of the form $D_{xw}$ etc. since propagators of this form are not part of the diagram we are
interested in. Proceeding with the first $\frac{\delta}{\delta J_a}$ we obtain,
\bea
&& D_{aw} \cdot DJ_x \cdot DJ_y \cdot DJ_z \cdot \left(JDJ\right)^4 + D_{ax} \cdot DJ_w \cdot DJ_y \cdot DJ_z \cdot 
\left(JDJ\right)^4 \nonumber \\
&&+ 8 \cdot DJ_w \cdot DJ_x \cdot DJ_y \cdot DJ_z \cdot 
\left(JDJ \right)^3 \cdot DJ_a
\eea
where we have differentiated $DJ_w$, $DJ_x$, and $(JDJ)^4$ in the first, second, and third expressions 
respectively. In the last expression, the 8 comes from acting on the $(JDJ)^4 $.  We proceed to the second 
$\frac{\delta}{\delta J_a}$. We obtain,
\bea
&&D_{aw} \cdot D_{ax} 			\cdot DJ_y \cdot DJ_z \cdot \left(JDJ\right)^4 \nonumber \\
&&+ 8 \cdot D_{aw} \cdot DJ_x \cdot DJ_y \cdot DJ_z \cdot \left(JDJ\right)^3 \cdot DJ_a\nonumber \\
&& + D_{aw} \cdot D_{ax}	\cdot DJ_y \cdot DJ_z\cdot\left(JDJ\right)^4 \nonumber \nonumber\\
&& + 8 \cdot D_{ax} \cdot DJ_w \cdot DJ_y \cdot DJ_z \cdot \left(JDJ\right)^3 \cdot DJ_a \nonumber\\
&& + 8 \cdot D_{aw} \cdot DJ_x \cdot DJ_y \cdot DJ_z \cdot \left(JDJ\right)^3 \cdot DJ_a \nonumber\\
&& +  8 \cdot D_{ax} \cdot DJ_w \cdot DJ_y \cdot DJ_z \cdot \left(JDJ\right)^3 \cdot DJ_a \nonumber\\
&& +  8 \cdot 6 \cdot DJ_w \cdot DJ_x \cdot DJ_y \cdot DJ_z \cdot 	\left(JDJ\right)^2 \cdot 
\left(DJ_a \right)^2
\eea 
and collecting like terms gives
\bea
&& 2 \cdot D_{aw} \cdot D_{ax} 		\cdot DJ_y \cdot DJ_z \cdot \left(JDJ\right)^4 \nonumber \\
&& 16 \cdot D_{aw} \cdot DJ_x \cdot DJ_y \cdot DJ_z \cdot \left(JDJ\right)^3 \cdot DJ_a
\nonumber\\
&& 16 \cdot D_{ax} \cdot DJ_w \cdot DJ_y \cdot DJ_z \cdot \left(JDJ\right)^3 \cdot DJ_a
\nonumber\\
&& 48 \cdot DJ_w \cdot DJ_x \cdot DJ_y \cdot DJ_z \cdot \left(JDJ\right)^2 \cdot \left(DJ_a \right)^2\nonumber\\
\eea 
There are two more $\frac{\delta}{\delta J_a}$'s.  The result is:
\bea
\left(2\cdot 8 		\cdot 6 + 16 \cdot 6 \cdot 2 + 16 \cdot 6 \cdot 2 + 48 \cdot 2 \right) D_{aw} \cdot D_{ax} 
\cdot DJ_y \cdot DJ_z \cdot \left(JDJ\right)^2 \cdot 			\left(DJ_a\right)^2
\eea
Combining, the coeffiecient obtained from the $a$ derivatives is 
\bea
\left(2\cdot 8 \cdot 6 + 16 \cdot 6 \cdot 2 + 16 \cdot 6 \cdot 2 + 48 \cdot 2 \right) = 12 \cdot 8 \cdot 6.
\eea
\\

One must proceed in the same fashion with the $b$, and $c$ derivatives. The result of this lengthy procedure is:
\bea
{\cal N}= 3! \cdot (16\cdot 14 \cdot 12 \cdot 10 ) \cdot (12\cdot 8\cdot 6) \cdot (24\cdot 4 \cdot 2) \cdot 24 \label{N}
\eea
Combining Eqns. (\ref{S}), (\ref{D}) and (\ref{N}) we obtain,
\bea
S= \frac{3! \cdot (16\cdot 14 \cdot 12 \cdot 10 ) \cdot (12\cdot 8\cdot 6) \cdot (24\cdot 4 \cdot 2) \cdot 24 }{3! \cdot (4!)^3 \cdot (2)^8 \cdot 8!}= 
\frac{1}{2} \label{ans1}
\eea
\\

The procedure described above, although straightforward, is incredibly tedious. The result can be obtained
immediately using the formula:
\bea
S = \frac{1}{R} \cdot \left(\frac{1}{2}\right)^{D_1} \cdot\left(\frac{1}{2!}\right)^{D_2} \cdot \left(\frac{1}{3!}\right)^{D_3} \cdot \left(\frac{1}{4!}\right)^{D_4}\label{form}
\eea
 where:  
 
 \xx $R$  =  the number of ways to permute the internal indices and produce an identical set of propagators
 
 \xx $D_1$  =  the number pairs of progatators of the form $D_{aa}^2$
 
 \xx $D_2$ = the numbers of pairs of propagators of the form $D_{mn}^2$ plus the number of factors of the form $D_{aa}^1$  (note: a factor $D_{aa}^2$ contributes 1 to $D_1$ and 2 to $D_2$ - these propagators are counted twice)
 
 \xx $D_3$  =  the number of triples of propagators of the form $D_{mn}^3$
 
 \xx $D_4$ = the number of sets of propagators of the form $D_{mn}^4$ \\
 
 We explain this notation by giving some examples.  Figs. [2a-2c] show three diagrams with non-unity values of $R$. We list the $R$ values and the indices or sets of indices that can be permuted without producing a different set of propagators. Figs. [3a-3d] contain examples of diagrams with different values of $D_1$, $D_2$, $D_3$ and $D_4$. The values are listed beneath each diagram. Finally, we note that the factor $(1/2)^{D_1}$ in (\ref{form}) is only different from one for the diagram in Fig. [4].\\

 For the diagram in Fig. [1], the values of the parameters defined above are: $R=1$, $D_1=D_3=D_4=0$ and $D_2=1$ (because there is a pair of propagators of the form $D_{bc}^2$).  
  Substitution into (\ref{form}) gives:
  \bea
  S= 1 \cdot \left(\frac{1}{2}\right)^0 \cdot \left(\frac{1}{2!}\right)^1 \cdot \left(\frac{1}{3!}\right)^0 \cdot \left(\frac{1}{4!}\right)^0 = \frac{1}{2} \label{ans2}
\eea 
in agreement with (\ref{ans1}). Some more complicated diagrams are shown in Fig. [5]. 

\section{Generalization to Other Theories}

The formula (\ref{form}) is valid for any diagram in $\phi^3$ theory, $\phi^4$ theory, spinor QED, scalar QED, or QCD. In this section we discuss the reason that the same formula works for different theories. It is not at all clear that this should be the case, since QED and QCD involve more than one type of field, including fields that are not self conjugate. Examples are the fermion fields in spinor QED $(\psi,\,\bar\psi)$ and the charged scalars in scalar QED $(\phi,\,\phi^\dagger)$.  To handle fields of this type the generating functional (\ref{gen1}) must be modified. We introduce separate sources for each field. A field that is not self conjugate is treated as two fields, and requires two sources (one for the field and one for its conjugate). For example, for QED the generating functional has the form (suppressing all indices, since they are not important when calculating symmetry factors):
\bea
W[J] = {\rm exp} \left\{ \int dx \!\int dy\left(\frac{i}{2}J_x D_{xy} J_y + i \bar K_x G_{xy}  K_y \right)+ i \int dx\,{\cal L}_{int}\left[
\frac{1}{i} \frac{\delta}{\delta J_x},\,\frac{1}{i} \frac{\delta}{\delta K_x},\,\frac{1}{i} \frac{\delta}{\delta \bar K_x}\right]\right\}\label{gen2}
\eea
where $K$ is the source for the field $\bar \psi$ or $\phi^\dagger$, $\bar K$ is the source for the field $\psi$ or $\phi$, and $J$ is the source for the photon field. The inverse photon propagator is $D(x-y)$ and the inverse fermion or scalar propagator is $G(x-y)$. 

\subsection{Propagators}

First we look at propagators. Consider what happens when we take functional derivatives of the generating functional (\ref{gen2}). Compare the two expressions:
\bea
&&\frac{\delta}{\delta J_y}\frac{\delta}{\delta J_x} \left(\frac{1}{2} J_z D_{zw}J_w\right) = \frac{1}{2}(D_{xy} + D_{yx}) = D_{xy} \nonumber \\
&& \frac{\delta}{\delta K_y}\frac{\delta}{\delta \bar K_x} \left(\bar K_z D_{zw}K_w\right) = G_{xy} 
\eea
Since the photon is self conjugate, we have $D_{xy}$ = $D_{yx}$. There is only one source function for the photon field and the factor of 1/2 in front of the photon inverse propagator is necessary to remove the factor of 2 that arises from taking the functional derivatives in either order. In contrast, the non-self-conjugate field has two independent degrees of freedom and thus we need two source functions, and no factor of 1/2 in front of the inverse propagator.  The propagator has a direction that corresponds to the flow of the associated quantum number and thus does not satisfy $G_{xy}=G_{yx}$. The point is that, for both types of fields, the generating functional is constructed so that the propagator appears without spurious factors of two.

\subsection{Vertices}

In order to understand the behaviour of vertices in QED and QCD we look at a few examples. Consider the interaction between two photons and two charged scalars in scalar QED. The interaction Lagrangian has the form,
\bea
{\cal L}_{int} = -e^2 g^{\mu\nu}\phi^\dagger A_\mu A_\nu \phi
\eea
The corresponding vertex can be obtained by functional differentiation of the generating functional in the usual way. The result is $\Gamma_{\mu\nu} = 2ie^2 g_{\mu\nu}$. The factor of two is a consequence of the fact that the two photon fields are interchangable. Now consider the calculation of the symmetry factor. Niavely we would write the vertex in functional derivative form as
\bea 
{\cal L}_{int} =  \left( \frac{\delta}{\delta J}\right)^2 \frac{\partial}{\delta K} \frac{\delta}{\delta \bar K}
\eea
Consider what happens if we proceed with the calcuation of the symmetry factor following the example for scalar fields described in section II. For a scalar theory, the factor of two which appears in the vertex (from the interchange of two fields), would have been included in the factor ${\cal N}$. Thus, if we want to treat the photons as scalars (for the purposes of calculating the symmetry factor) we have to include an extra factor 1/2 in the interaction. The `scalarized' interaction has the form:
\bea 
{\cal L}_{int} =  \frac{1}{2}\left( \frac{\delta}{\delta J}\right)^2 \frac{\partial}{\delta K} \frac{\delta}{\delta \bar K}
\eea

Now, consider the three gluon vertex in QCD. This vertex is obtained from the part of the interaction Lagrangian that contains three gluon fields which can be written:
\bea
{\cal L}_{int} = -g f_{abc} \partial_\mu A_\nu ^a A^\mu_b A^\nu_c
\eea
The corresponding vertex is:
\bea
\Gamma_{\mu\nu\rho}^{abc}(p,q,r) = g f_{abc}[g_{\mu\nu}(p-q)_\rho + g_{\nu\rho}(q-r)_\mu + g_{\rho\mu}(r-p)_\nu]
\eea
 Note that the vertex contains six terms. In a scalar theory, these six terms would be identical and would give rise to a factor of six, were it not for the fact that the interaction Lagrangian contains an explicit factor of $1/3!$ which exactly cancels this factor of six. Therefore, to obtain the symmetry factor, we can use a `scalarized' interaction Lagrangian of the form
\bea
{\cal L}_{int} = \frac{1}{3!}\left( \frac{\delta}{\delta J}\right)^3
\eea
where $J$ is the source for the gluon field, and all indices are ignored.  The point is that this interaction has exactly the same form as the interaction for $\phi^3$ theory considered above.

The four gluon vertex works the same way: The interaction Lagrangian is 
\bea 
{\cal L}_{int} = -\frac{g^2}{4}  f_{abc} f_{ade} g_{\mu\nu} g_{\lambda\tau} A^\mu_b A^\nu_c A^\lambda_d A^\tau _e
\eea
The Feynman rule for the vertex contains six terms. Combining the factor of 1/4 in the interaction Lagrangian and a factor of 1/6 so that we can treat the gluons as scalars, we need a factor of $24=4!$ in the `scalarized' form of the interaction vertex:
\bea
{\cal L}_{int} = \frac{1}{4!}\left( \frac{\delta}{\delta J}\right)^4
\eea
This interaction has the same form as the interaction for $\phi^4$ theory considered above.\\

All interactions in spinor and scalar QED and QCD can be written in a `scalarized' form by including a factor 
\bea
\Pi_i \frac{1}{N_i!}
\eea
where $N_i$ is the number of fields of type $i$ in the vertex. 

\subsection{Symmetry factors}

Using the discussion in the sections above about the behaviour of propagators and vertices for non-scalar theories, we can consider taking functional derivatives to construct a given diagram, following the method outlined in section II. It is straightforward to see that symmetry factors are given by the same formula as for scalar theory (\ref{form}) by using the rules below (in all figures, solid lines are uncharged scalars, solid lines with arrows are fermions or charged scalars (or ghosts), and dotted lines are photons or gluons):\\

\xx a) Factors of $G$ make no contribution to the factors $D_1$, $D_2$, $D_3$, $D_4$ in (\ref{form}). For example, the first diagram in Fig. [6] has $D_2=1$ and thus $S=1/2$, while the second diagram in Fig. [6] has $D_2=0$ and $S=1$.

\xx b) The fact that $G_{xy} \ne G_{yx}$ must be taken into account when calculating $R$ factors. For example, the first diagram in Fig. [7] has $R=2$, but the second diagram in Fig. [7] has $R=1$. \\

\xx Fig. [8] gives some more examples of diagrams for which the symmetry factor can be calculated using (\ref{form}).

\section{Conclusion}
The calculation of symmetry factors is a straightforward but tedious business. The formula derived in this paper (\ref{form}) provides a simple way to obtain the symmetry factor for any diagram in $\phi^3$ theory, $\phi^4$ theory, spinor QED, scalar QED, or QCD.

\begin{figure}
\epsfxsize=4cm
\epsfysize=2cm
\centerline{\epsfbox{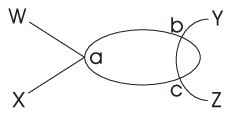}}
\vskip 0.4cm
\centerline{{\scriptsize Fig.~[1]:  Diagram from $\phi^4$ theory}}
 \label{F1a}
 \end{figure}

\begin{figure}
\epsfxsize=13cm
\epsfysize=2.5cm
\centerline{\epsfbox{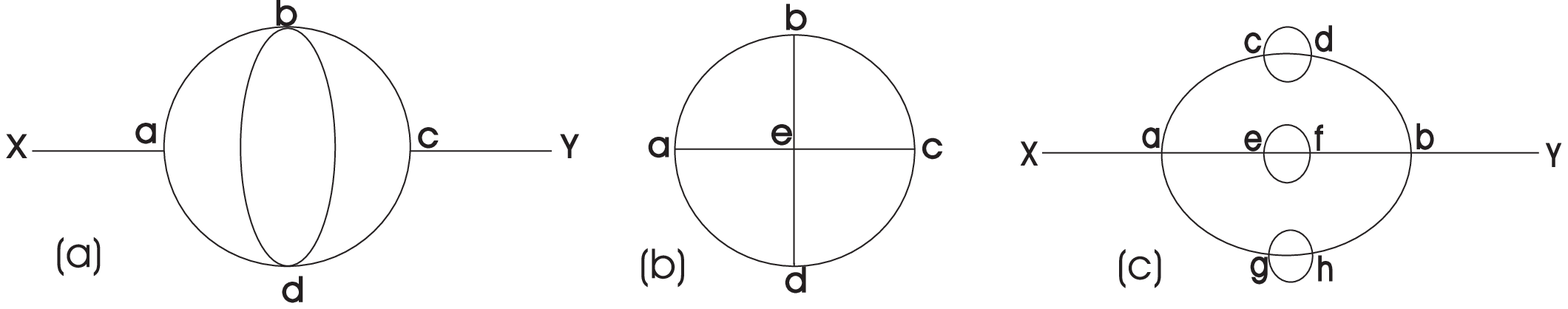}}
\vskip 0.4cm
\centerline{{\scriptsize Fig.~[2]:  a) $R=2$: $\{b, d\}$; b) $R=4$: $\{b, d\}$ and $\{a,c\}$; c) $R=3!=6$: $\{(c,d),(e,f),(g,h)\}$}}
 \label{F2}
 \end{figure}

\begin{figure}
\epsfxsize=12cm
\epsfysize=2cm
\centerline{\epsfbox{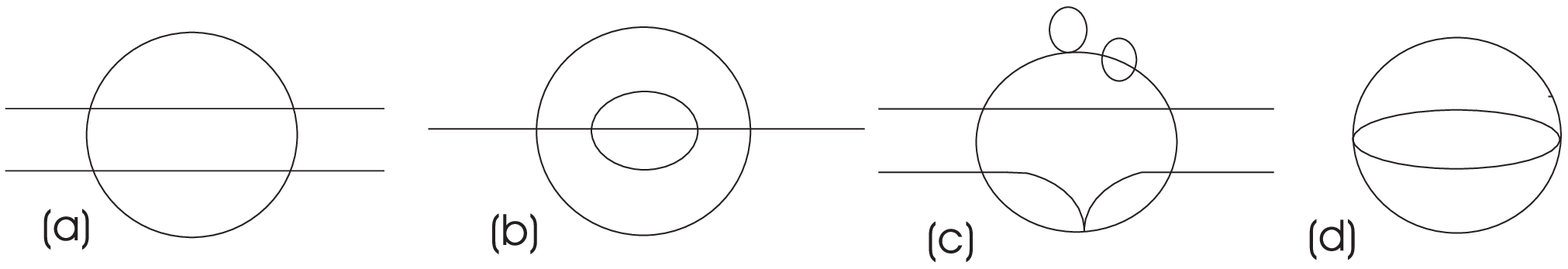}}
\vskip 0.4cm
\centerline{{\scriptsize Fig.~[3]:  a) $R=1$, $D_1=0$, $D_2=2$, $D_3=0$, $D_4=0$; b) $R=1$, $D_1=0$, $D_2=1$, $D_3=1$, $D_4=0$}}
\centerline{{\scriptsize c) $R=1$, $D_1=0$, $D_2=3$, $D_3=1$, $D_4=0$; d) $R=2$, $D_1=0$, $D_2=0$, $D_3=0$, $D_4=1$}}
 \label{F1}
 \end{figure}
   
\begin{figure}
\epsfxsize=4cm
\epsfysize=1.5cm
\centerline{\epsfbox{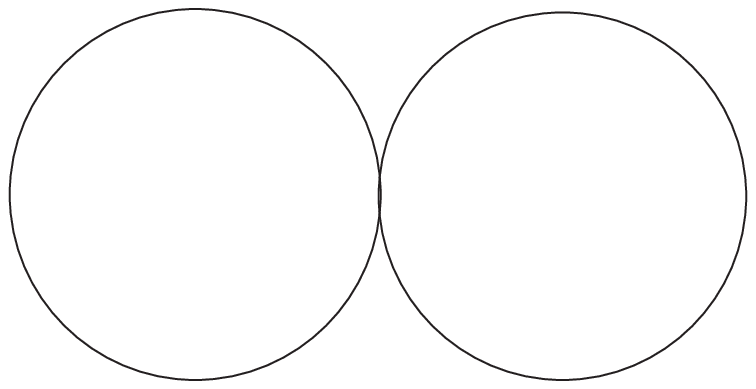}}
\vskip 0.4cm
\centerline{{\scriptsize Fig.~[4]:  $R=1$, $D_1=1$, $D_2=2$, $D_3=0$, $D_4=0$}}
 \label{F4}
 \end{figure}

\begin{figure}
\epsfxsize=12cm
\epsfysize=1.8cm
\centerline{\epsfbox{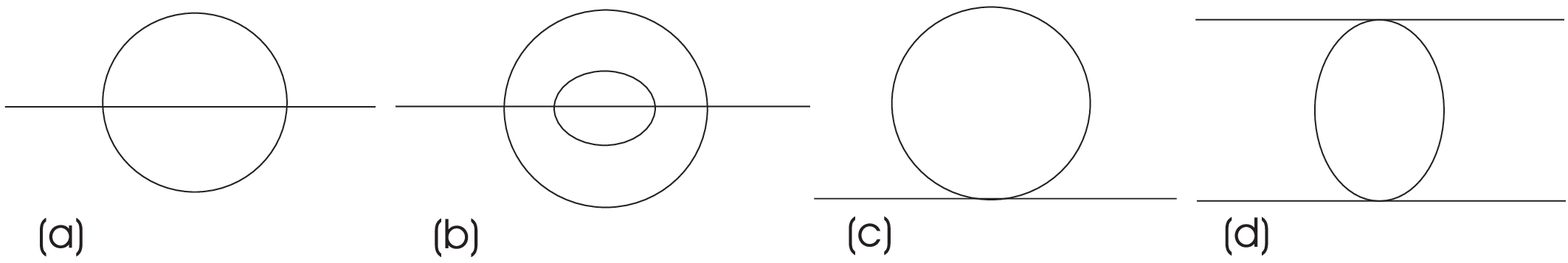}}
\vskip 0.4cm
 \label{F4}
 \end{figure}
 \begin{figure}
\epsfxsize=14cm
\epsfysize=1.8cm
\centerline{\epsfbox{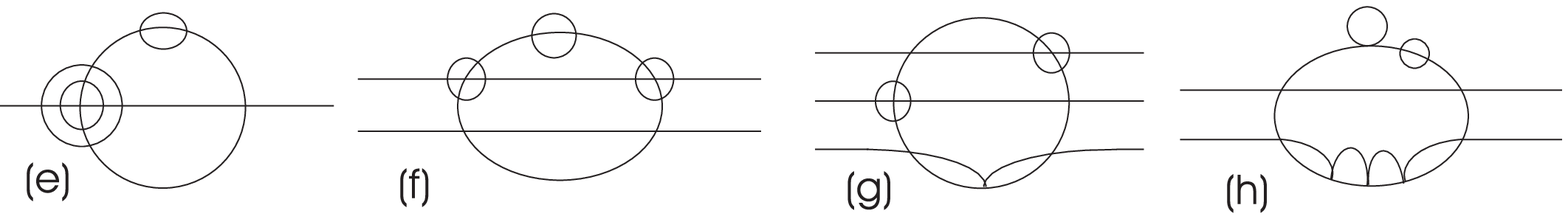}}
\vskip 0.4cm
\centerline{{\scriptsize Fig.~[5]: Some diagrams for scalar theories.}}
\centerline{{\scriptsize Symmetry factors are: a) 1/6; b) 1/12; c) 1/2; d)
1/2; e) 1/6; f) 1/12; g) 1/4; h) 1/192 }}
 \label{F5}
 \end{figure}




 \begin{figure}
\epsfxsize=9cm
\epsfysize=2cm
\centerline{\epsfbox{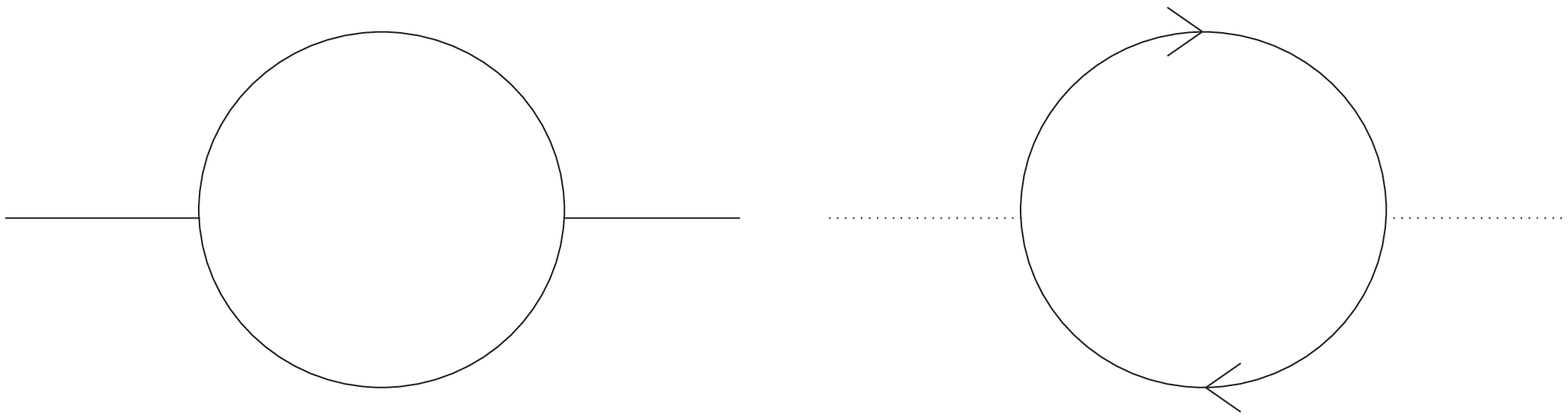}}
\vskip 0.4cm
\centerline{{\scriptsize Fig. [6]: D factors for theories with non-self-conjugate fields}}
 \label{F6}
 \end{figure}
 \begin{figure}
\epsfxsize=9cm
\epsfysize=2cm
\centerline{\epsfbox{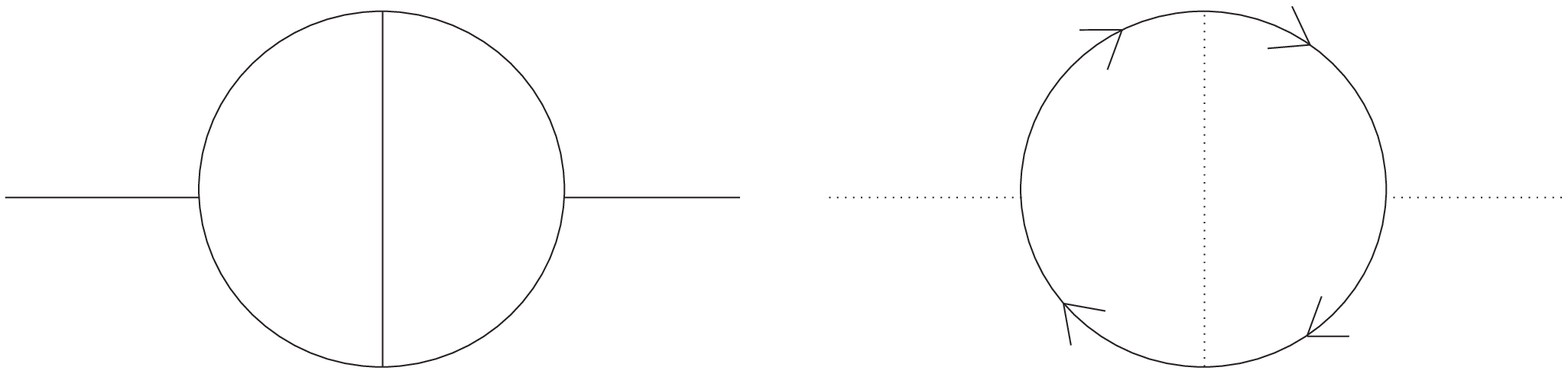}}
\vskip 0.4cm
\centerline{{\scriptsize Fig. [7]: R factors for theories with non-self-conjugate fields}}
 \label{F7}
 \end{figure} 
\begin{figure}
\epsfxsize=11cm
\epsfysize=2cm
\centerline{\epsfbox{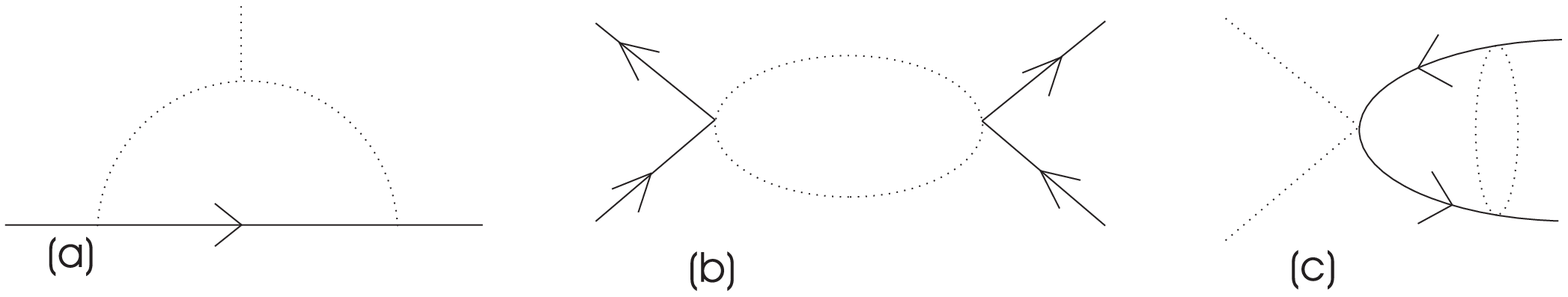}}
\vskip 0.4cm
 \label{F5a}
 \end{figure}
\begin{figure}
\epsfxsize=9.5cm
\epsfysize=2cm
\centerline{\epsfbox{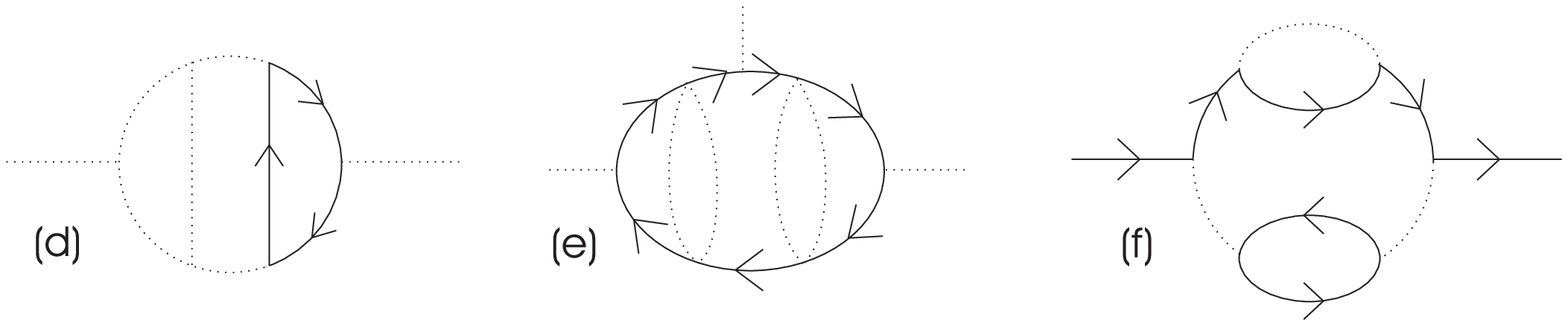}}
\vskip 0.4cm
\centerline{{\scriptsize Fig. [8]: Some diagrams with non-self-conjugate fields.}}
\centerline{{\scriptsize Symmetry factors are: a) 1; b) 1/2; c)
1/2; d) 1; e) 1/4; f) 1 }}
 \label{F8}
 \end{figure}




\begin{thebibliography}{}

\bibitem {bb} See for example: `Quantum Field Theory,' C. Itzykson and J.B. Zuber, McGraw Hill (New York, 1980); `An Introduction to Quantum Field Theory,' M.E. Peskin and D.V. Schroeder, Perseus Books (Cambridge, Massachusetts, 1995); `Quantum and Statistical Field Theory,' M. Le Bellac, Oxford University Press (Oxford, 1997).

\end{thebibliography}
\end{document}